\documentclass{article}
\usepackage{amsmath,amssymb,amsthm,bm,helvet}
\usepackage[mathscr]{eucal}
\usepackage{graphicx}

\newcommand{\ve}[1]{\bm{#1}} 
\newcommand\tit[1]{``#1,''}

\newcommand{\ie}{{\it i.e.},}
\newcommand{\eg}{{\it e.g.},}
\newcommand{\Eq}[1]{Eq.~(\ref{#1})}
\newcommand{\Sec}[1]{Sec.~\ref{Sec:#1}}
\newcommand{\Fig}[1]{Fig.~\ref{Fig:#1}}

\newcommand{\FI}[1]{\mathcal{I}_{#1}}

\newcommand{\DI}[1]{\mathcal{D}_{#1}}

\newcommand{\JI}[2]{\mathscr{J}_{#1}^{#2}}

\newcommand{\EqI}[1]{\cite[Eq.~#1]{SCLD1}}

\newcommand{\cld}{\mu}
\newcommand{\seg}{\iota}
\newcommand{\dst}{\eta}
\newcommand{\cor}{\gamma}

\newcommand{\av}[1]{\langle{#1}\rangle}
\newcommand{\CB}{\mathfrak{V}} 
\newcommand{\NB}{\mathfrak{N}} 
\newcommand{\B}{\mathfrak{B}} 

\newcommand{\R}{\mathbb{R}}
\newcommand{\Sph}{\mathbb{S}}
\newcommand{\T}{\mathcal{T}}

\newcommand{\Rd}{\mathsf R}
\newcommand{\Lh}{\mathsf L}

\newcommand{\TRI}{\blacktriangle}
\newcommand{\SQU}{\blacksquare}
\newcommand{\LLint}[1]{(\Lh\cap{#1})_\rightthreetimes^2}

\newcommand{\sigseg}{\seg_\pm}
\newcommand{\sigcld}{\cld_\pm}

\newcommand{\intd}[1]{\int\!\!\!\!\mathop{\mathstrut~#1}\limits}

\newcommand{\w}[1]{{\breve{#1}}}

\numberwithin{equation}{section}

\newcommand{\scrf}{\fontfamily{cmfr}\selectfont}

\title{\sf\bfseries Signed Chord Length Distribution\\
 Part II}
\date{}
\author{\sf\itshape Alexander \t{Yu}.~Vlasov}
\begin{document}
\sloppy

\maketitle

\makeatletter
\global\@specialpagefalse
\renewcommand{\@oddfoot}{\rlap{\copyright{\it\scrf
\ A.\t{Yu}.Vlasov, 2009}} \hfil \thepage \hfil \llap{\sf Signed CLD II}}

\renewcommand\section{\@startsection {section}{1}{\z@}%
                                   {-3.5ex \@plus -1ex \@minus -.2ex}%
                                   {2.3ex \@plus.2ex}%
                                   {\normalfont\large\sf\bfseries\boldmath\uppercase}}
\renewcommand\subsection{\@startsection{subsection}{2}{\z@}%
                                     {-3.25ex\@plus -1ex \@minus -.2ex}%
                                     {1.5ex \@plus .2ex}%
                                     {\normalfont\sf\bfseries\itshape\boldmath}}
\makeatother

\begin{abstract}
 This paper continues description of applications of {\em signed 
chord length distribution} started in \cite{SCLD1}, arXiv:0711.4734 [math-ph]. 
It is shown simple relation between equation for some transfer 
integrals with source and target bodies and different geometrical
distributions for union of this bodies. The union of disjoint 
bodies is always nonconvex object and for such a case derivatives
of correlation function (used for definition of signed radii 
and chord lengths distributions) always produce (quasi)densities with 
negative values. Many equations used in this part are direct 
consequences of analogue formulas in \cite{SCLD1}.
\end{abstract}

{\sf\boldmath\tableofcontents}

\section{Introduction to Part II}
\label{Sec:IIntro}

\begin{figure}[hbt]
\begin{center}
\includegraphics[scale=0.75]{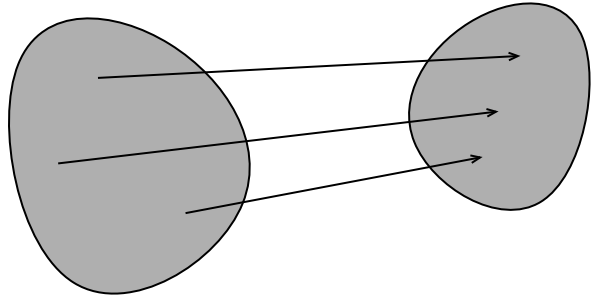}
\end{center}
\caption{Scheme of integration on ``source'' and ``target'' bodies}
\label{Fig:transf12}
\end{figure}

In first part \cite{SCLD1} was considered a basic theory
of signed chord length distribution. Here is discussed an
extension for specific case. Let us consider two bodies $\CB_1$, $\CB_2$
(see \Fig{transf12}) with volumes $V_1$, $V_2$ and integral
\begin{equation}
\JI{\CB_1}{\CB_2}(\Phi) = 
 \int_{\CB_1}\int_{\CB_2}\Phi(R)\, d\ve{r}\, d\ve{r}' =
 \JI{\CB_2}{\CB_1}(\Phi),
\quad R = |\ve{r}-\ve{r}'|,
\quad \JI{\CB_1}{\CB_2}(1) = V_1 V_2, 
\label{Int12}
\end{equation}
where $\ve{r} \in \CB_1$, $\ve{r}' \in \CB_2$, 
and $d\ve{r}$, $d\ve{r}'$ are two 
three-dimensional volume elements.

A particular case is two equivalent bodies $\CB_1 = \CB_2 = \CB$ and 
integral \EqI{A1} $\FI{\CB}(\Phi) \equiv \JI{\CB}{\CB}(\Phi)/V^2$. 
On the other hand, for two different disjoint bodies 
it is possible to consider union $\NB = \CB_1 \cup \CB_2$ as a single
compound nonconvex object and due to simple decomposition
\begin{equation}
\intd{d\ve{r}}_{\CB_1 \cup \CB_2}\intd{d\ve{r}'}_{\CB_1 \cup \CB_2}
\!\!\!\Phi(R)  = 
\!\intd{d\ve{r}}_{\CB_1}\!\intd{d\ve{r}'}_{\CB_1} \Phi(R) +
\!\intd{d\ve{r}}_{\CB_1}\!\intd{d\ve{r}'}_{\CB_2} \Phi(R) +
\!\intd{d\ve{r}}_{\CB_2}\!\intd{d\ve{r}'}_{\CB_1} \Phi(R) +
\!\intd{d\ve{r}}_{\CB_2}\!\intd{d\ve{r}'}_{\CB_2} \Phi(R)
\label{sumiint}
\end{equation}
it is possible to express \Eq{Int12} using single-body integrals $\FI{}$
from \cite{SCLD1}
\begin{equation}
 2 \, \JI{\CB_1}{\CB_2}(\Phi) =  (V_1+V_2)^2 \FI{\CB_1 \cup \CB_2}(\Phi) 
 - V_1^2 \FI{\CB_1}(\Phi) -  V_2^2\FI{\CB_2}(\Phi).
\label{I1U2}
\end{equation}

This consideration justifies application of signed chords and radii distribution
for calculation of ``transfer integrals'' like \Eq{Int12}.  It is shown
below, that (quasi)density functions for signed radii and chord length distributions 
introduced in \cite{SCLD1} for \Eq{Int12} with nonoverlapping bodies are always 
have negative values, even if all three terms in 
\Eq{I1U2} are convex.

\begin{figure}[hbt]
\begin{center}
\includegraphics[scale=0.75]{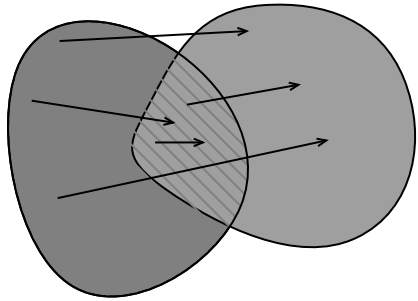}
\end{center}
\caption{Two overlapping bodies}\label{Fig:overlap}
\end{figure}

It is also possible to consider two overlapping bodies $\CB_1$ and $\CB_2$ 
\Fig{overlap}. Such a case may be described using three nonoverlapping 
bodies: $\B_3 = \CB_1 \cap \CB_2$, $\B_1 = \CB_1\setminus\B_3$, 
$\B_2 = \CB_2\setminus\B_3$, {\ie} $\CB_1 = \B_1 \cup \B_3$, 
$\CB_2 = \B_2 \cup \B_3$ and equations for nonoverlaping or equal bodies
\begin{equation}
 \JI{\CB_1}{\CB_2} = \JI{\B_1 \cup \B_3}{\B_2 \cup \B_3} =
 \JI{\B_1}{\B_2} + \JI{\B_1}{\B_3} + \JI{\B_3}{\B_2} + \JI{\B_3}{\B_3}.
\label{Bover}
\end{equation}

Such a scheme lets consider only disjoint bodies without lost of generality.
Presentation suggests an acquaintance with first part \cite{SCLD1} and
may be considered as extension of corresponding sections. \Sec{CorDst}
develops methods discussed in \cite[App.~A-1, A-2]{SCLD1}. \Sec{SigMx} is
relevant with \cite[Sec.~3]{SCLD1} and generalizations
briefly discussed in \Sec{Nonun}, \ref{Sec:Paths} are analogues 
of \cite[Sec.~4,~5]{SCLD1}.

\section{Distances and correlations}
\label{Sec:CorDst}

\subsection{Distribution of distances}
\label{Sec:DisDst}

Let's write generalization of formula \EqI{A2} for distribultion 
of distances $\dst_{12}$ between points in $\CB_1$ and $\CB_2$ 
\begin{equation}
 \frac{1}{V_1 V_2}\JI{\CB_1}{\CB_2}(\Phi) \equiv 
 \frac{1}{V_1 V_2}\int_{\CB_1}\int_{\CB_2}\Phi(|\ve{r}-\ve{r}'|)\, d\ve{r}\, d\ve{r}'
= \int_0^\infty {\Phi(x) \dst_{12}(x) dx},
\label{IntV1V2dst}
\end{equation}
where $1/(V_1 V_2)$ is multiplier  used for normalization 
\(\int_0^\infty \dst_{12}(x) dx  = 1.\) 
Proof of \Eq{IntV1V2dst} is analogue of {\em Lemma~1} in \cite[Appendix A-1]{SCLD1}.

From equations for union like \Eq{sumiint} or \Eq{I1U2} may be derived similar
expression for distributions of distances
\begin{equation}
 (V_1+V_2)^2 \dst_{1 \cup 2}(l) =  
 V_1^2 \dst_1(l) + 2 V_1 V_2 \dst_{12}(l) + V_2^2 \dst_2(l),
\label{dst1U2}
\end{equation}
where $\dst_1$, $\dst_2$ and $\dst_{1 \cup 2}$ correspond to definition
of distances distribution for single (convex or nonconvex) body used in first part,
{\em Definition 1} \cite{SCLD1}. There is also reason to generalize such
equation for arbitrary number of bodies and write
\begin{equation}
 \Bigl(\sum_{k=1}^n V_k\Bigr)^2 \dst_\cup(l) 
 = \sum_{k=1}^n V_k^2 \dst_k(l) + 2\!\sum_{j=i+1}^n \sum_{i=1}^n V_i V_j\, \dst_{ij}(l)
 = \sum_{i,j=1}^n V_i V_j\, \dst_{ij}(l),
\label{dstU}
\end{equation}
where $\dst_\cup$ is distribution of distances for union of $n$ bodies 
(considered as a single object) and notation $\dst_{ii}=\dst_i$ is 
indirectly used for convenience in last expression 
with single sum $\sum_{ij}$. Such notation make possible to talk about
$\dst_{ij}$ as about some {\em matrix-valued density\/} $\bm{\dst}(l)$.

\subsection{Correlation function}
\label{Sec:CorFun}

Correlation function $\cor(\ve{r})$, $\ve{r} \in \R^3$ or $\cor(l)$, $l \in \R$
may be defined for two densities $\rho_1(\ve{r})$, $\rho_2(\ve{r})$ 
$\ve{r} \in \R^3$ as ({\em cf\/} \EqI{A5})
\begin{equation}
 \cor_{12}(\ve{r}) = \int_{\R^3} \rho_1(\ve{r}') \rho_2(\ve{r}+\ve{r}') d \ve{r}', \quad
 \cor_{12}(l) = \frac{1}{4\pi l^2}\int_{\Sph_l} \cor_{12}(\ve{r}) d\Omega, \quad 
 d\Omega = \sin\theta\, d\theta\, d\phi,
\label{corr2}
\end{equation}
{\ie} $\cor_{12}(l)$ is an average of $\cor_{12}(\ve{r})$ on sphere with radius $l$,
$\{\Sph_l : |\ve{r}|=l \}$. 

In simplest case of two bodies $\CB_1$ and $\CB_2$ with constant unit density 
$\rho_k(\ve{r}) = 1$ for $\ve{r} \in \CB_k$ and zero otherwise.
It is possible to rewrite \Eq{Int12}
\begin{subequations}\label{Int12C4}
\begin{eqnarray}
\JI{\CB_1}{\CB_2}(\Phi) &=&  
 \int_{\R^3} \int_{\R^3} \rho_1(\ve{r}) \rho_2(\ve{r}')%
 \Phi\bigl(|\ve{r}'-\ve{r}|\bigr) d\ve{r}\, d\ve{r}' \\
 &=& \int_{\R^3} \int_{\R^3} \rho_1(\ve{r}) \rho_2(\ve{r+R})%
 \Phi\bigl(|\ve{R}|\bigr) d\ve{r}\, d\ve{R} \qquad (\ve{R=r'-r}) \\
 &=& \int_{\R^3} \cor_{12}(\ve{R})\Phi\bigl(|\ve{R}|\bigr) d\ve{R} 
\label{Int12corr} \\
 &=& 4\pi \int_0^\infty l^2 \cor_{12}(l)\Phi(l) dl,
\label{Int12cor}
\end{eqnarray}
\end{subequations}
where \Eq{Int12cor} is produced from \Eq{Int12corr} by integration 
over spheres $\Sph_l$. It may be compared with analogue integrals
for autocorrelation function \EqI{A6, A7} up to constant multiplier $1/V$, 
because here {\em is not used} normalization multiplier $V^{-1/2}$ 
for density introduced in \cite{SCLD1}.

Comparison of \Eq{Int12cor} and \Eq{IntV1V2dst} with arbitrary function $\Phi(l)$
produces relation between $\cor_{12}(l)$ and $\dst_{12}(l)$
\begin{equation}
\dst_{12}(l) = \frac{4 \pi}{V_1 V_2} l^2 \cor_{12}(l).
\label{cor2dst}
\end{equation}

Due to \Eq{dst1U2} and \Eq{cor2dst}
\begin{equation}
 \cor_{1 \cup 2}(l) =  \cor_{11}(l) + 2\cor_{12}(l) + \cor_{22}(l),
\label{cor1U2}
\end{equation}
where $\cor_{1 \cup 2}$, $\cor_{11}$ and $\cor_{22}$ are autocorrelation 
function without normalization, {\ie} $\cor_{11}(0) = V_1$, $\cor_{22}(0) = V_2$
and $\cor_{1 \cup 2}(0) = V_1 + V_2$ (for nonoverlapping bodies).
Here is more convenient to do not use normalization {\em vs} \cite{SCLD1}
to make expressions like \Eq{cor1U2} more clear. 

For example, due to \Eq{dstU} and \Eq{cor2dst} there is quite simple expression 
with (auto)\-cor\-re\-la\-tion functions for few objects 
\begin{equation}
 \cor_\cup(l)  = \sum_{k=1}^n \cor_{kk}(l) + 2 \sum_{j=i+1}^n \sum_{i=1}^n \cor_{ij}(l)
 = \sum_{i,j=1}^n \cor_{ij}(l),
\label{corU}
\end{equation}
Here again appears some {\em matrix} $\bm{\cor}(l)$ with correlation functions 
$\cor_{ij}$. 

\section{Signed matrix-valued distributions}
\label{Sec:SigMx}

\subsection{Integration by parts}
\label{Sec:IntPart}

Similarly with \EqI{3.1} it is possible to write for \Eq{Int12} with
$\Phi(x) = \varphi(x)/(4 \pi x^2)$ due to \Eq{IntV1V2dst}, \Eq{cor2dst}
and usual formula for integration by parts
\begin{subequations}\label{Int12B4}
\begin{eqnarray}
\JI{\B_1}{\B_2}(\Phi) \equiv
\int_{\B_1} \int_{\B_2}\frac{\varphi(R)}{4 \pi R^2} d\ve{r}\, d\ve{r}' 
 &=& V_1 V_2\int_0^\infty\!\!\frac{\varphi(x)}{4 \pi x^2}\dst_{12}(x)dx \label{Int12DstB} \\
 &=& \int_0^\infty\!\!\cor_{12}(x)\varphi(x)dx \label{Int12CorB} \\
 &=& -\int_0^\infty\!\!\cor'_{12}(x)%
 \left(\int_0^x\!\!\varphi(r)dr \right) dx \label{Int12CorB'} \\
 &=& \int_0^\infty\!\!\cor''_{12}(x)%
 \left(\int_0^x\!\!\!\int_0^p\!\!\varphi(r)dr\,dp\right) dx, \label{Int12CorB''}
\end{eqnarray}
\end{subequations}
where $R=|\ve{r}-\ve{r}'|$.
Here $\B_1$ and $\B_2$ may be without lost of generality considered
nonoverlapping due to adaptability of decompositions like 
\Fig{overlap} and \Eq{Bover} discussed earlier, \Eq{Int12DstB} 
and \Eq{Int12CorB} follows from \Eq{IntV1V2dst} and \Eq{cor2dst} respectively,
but \Eq{Int12CorB'} and \Eq{Int12CorB''} are produced by formal
integrations by parts and need for further explanation.

Comparison with definition of Dirac integral used in \cite{SCLD1}
\begin{equation}
\DI{\B}(\varphi) \equiv \frac{1}{V}\int_\B \int_\B%
 \frac{\varphi(|\ve{r}-\ve{r}'|)}{4 \pi |\ve{r}-\ve{r}'|^2} d\ve{r}\, d\ve{r}'
\label{I-IntDir}
\end{equation}
produces links with considered integrals, if to choose $\B = \B_1 \cup \B_2$
and use equations like \Eq{sumiint} and \Eq{I1U2}
\begin{equation}
(V_1+V_2)\DI{\B_1 \cup \B_2}(\varphi) = 
\JI{\B_1 \cup \B_2}{\B_1 \cup \B_2}(\Phi) =
2 \JI{\B_1}{\B_2}(\Phi) + V_1 \DI{\B_1}(\varphi) + V_2 \DI{\B_2}(\varphi)
\label{DIJI12}
\end{equation}

\subsection{Some properties of distributions for two bodies}
\label{Sec:Dist2}

Signed chords $\sigcld(l)$ and radii $\sigseg(l)$ distributions was
defined in first part \cite{SCLD1} via formulas 
\begin{equation}
 \sigseg(l) = -\cor'(l), 
 \quad \sigcld(l) = \av{l} \cor''(l), \quad
 \av{l} = \int_0^\infty \!\!\!\! l \sigcld(l)\,dl 
= \frac{1}{\cor'(0)} = \frac{4V}{S},
\label{sigsc}
\end{equation}
where $\cor(l)$ is (normalized) autocorrelation function, $S$ and $V$
are surface area and volume of given body utilized due to
{\em Cauchy relation for average chord length} $\av{l} = 4V/S$. 

\begin{figure}[hbt]
\begin{center}
\includegraphics[scale=0.24]{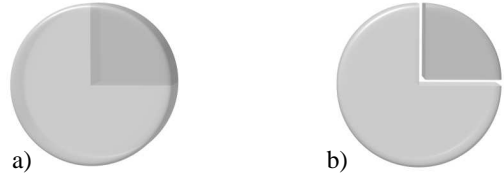}
\end{center}
\caption{a) Common boundaries. b) Displacement}\label{Fig:overbnd}
\end{figure}

Distribution of distances between two bodies $\dst_{12}$ in \Eq{IntV1V2dst} 
and \Eq{Int12DstB} has clear geometrical meaning.
Integrals \Eq{Int12CorB'} and \Eq{Int12CorB''} makes reasonable to introduce
analogues of \Eq{sigsc}
\begin{equation}
 C_{12}^\seg\, \seg_{12}(l) = - \cor'_{12}(l), 
\quad
 C_{12}^\cld\, \cld_{12}(l) =  \cor''_{12}(l), 
\label{Cscld12}
\end{equation}
where $C_{12}^\seg$ and $C_{12}^\cld$ are some constants.
It is useful to apply \Eq{Int12B4}, \Eq{DIJI12} and \Eq{sigsc}
in analogues of \Eq{dst1U2} for disjoint bodies  
\begin{equation}
 (V_1+V_2)\, \sigseg^{\B_1 \cup \B_2}(l) =  V_1\, \sigseg^{\B_1}(l) 
  +V_2\, \sigseg^{\B_2}(l) + 2 C_{12}^\seg\, \seg_{12}(l)
\label{sum12seg}
\end{equation}
and due to {\em Cauchy relation} for $\B_1$ and $\B_2$ with 
surface areas $S_1$ and $S_2$
\begin{equation}
 (S_1 + S_2)\,\sigcld^{\B_1 \cup \B_2}(l) 
 =  S_1\,\sigcld^{\B_1}(l) 
  + S_2\,\sigcld^{\B_2}(l) + 2 C_{12}^\cld\, \cld_{12}(l).
\label{sum12cld}
\end{equation}
Here bodies with common parts of boundaries \Fig{overbnd}a and total area 
$S_{12} < S_1 + S_2$ are excluded for simplicity, but may
be considered using infinitesimal displacement of overlapped 
surfaces \Fig{overbnd}b.

The distributions of distances $\dst_1$, $\dst_2$, $\dst_\cup$ and even 
$\dst_{12}$ in \Eq{dst1U2} are traditional density functions with simple 
geometrical and statistical interpretation as distribution of distances
between points in single object or two different bodies.
For convex body radii and chord density functions $\seg$ and $\cld$
also have clear meaning. In first part \cite{SCLD1} was represented
interpretation of signed radii and chord length (quasi)density functions 
for nonconvex body $\sigseg$ and $\sigcld$ via composition of some density 
functions with alternating signs. 

Similar decompositions for $\seg_{12}$ and $\cld_{12}$ are represented
below, but there is additional problem with definition of 
$C_{12}^\seg$ and $C_{12}^\cld$ due to impossibility to use idea
of unit normalization for distribution, because integrals over
$\seg_{12}$ and $\cld_{12}$ for two nonoverlapping bodies are zeros. 
It is clear already from integration of \Eq{sum12seg} and \Eq{sum12cld}.

{\em So, such functions always must have both positive and negative values}
and in considered approach analogue of Dirac integrals may {\em never} use
expression with probability density function satisfying usual definition 
with nonnegativity and unit normalization conditions. Both $\seg_{12}(l)$ 
and $\cld_{12}(l)$ are reasonable examples of  ``Feynman's negative
probabilities'' discussed in first part \cite{SCLD1}.

It may be simpler to consider (signed) density functions as elements
of some {\em transition matrices}. Let us use distributions of distances
as simple example. In \Sec{DisDst} was introduced matrix $\bm{\dst}$
with all elements $\dst_{ij}(l)$ are density functions with unit 
integral $\int_0^\infty \dst_{ij} (l) dl = 1$.

On the other hand it is possible to consider only $\dst_\cup$ as density
function with property $\int_0^\infty \dst_\cup (l) dl = 1$ and introduce
matrix $\bm{\w{\dst}}$ with components
\begin{equation}
\w{\dst}_{ij}(l) = \frac{V_i V_j}{V^2_\cup}\dst_{ij}(l), \quad  
 V_\cup = \sum_{k=1}^n V_k\Bigr.
\label{wdst}
\end{equation}
Instead of \Eq{dstU} it is possible to write
\begin{equation}
\dst_\cup(l) = \sum_{i,j=1}^n \w{\dst}_{ij}(l),
\label{dstW}
\end{equation}
{\ie} $\dst_\cup$ is density function for distribution of distances and 
\Eq{dstW} is a sum of contributions $\w{\dst}_{ij}$ for $n^2$ 
possible combinations with different pairs of bodies. 

Here due to \Eq{wdst}
\begin{equation}
 p_{ij} = \int_0^\infty\!\!\! \w{\dst}_{ij}(l)\, dl = \frac{V_i V_j}{V^2_\cup}
\label{dstpij}
\end{equation}
is probability for first and second points to lay in $\B_i$ and $\B_j$
respectively. 

\smallskip

Let's introduce a similar matrix $\bm{\w{\seg}}$ for assembly with few 
{\em disjoint} bodies as
\begin{equation}
\w{\seg}_{ij}(l) = -\frac{1}{V_\cup}\cor'_{ij}(l).
\label{wseg}
\end{equation}
For particular case $i=j$ 
\begin{equation}
\w{\seg}_{ii}(l) = \frac{V_i}{V_\cup}\sigseg^{\B_i}(l)
\label{wsegii}
\end{equation}
Due to such definition and \Eq{corU} signed radii distribution for union
of such bodies may be expressed as
\begin{equation}
 \sigseg(l) = \sum_{i,j=1}^n \w{\seg}_{ij}(l).
\label{segW}
\end{equation}
It is an analogue of \Eq{dstW}. 

Now it is possible rewrite \Eq{Int12CorB'} for a pair in such collection 
\begin{equation}
\frac{1}{V_\cup}\int_{\B_i}\int_{\B_j}%
 \frac{\varphi(|\ve{r}-\ve{r}'|)}{4 \pi |\ve{r}-\ve{r}'|^2} d\ve{r}\, d\ve{r}' 
 =  \int_0^\infty\!\!\w{\seg}_{ij}(x)%
 \left(\int_0^x\!\!\varphi(r)dr \right) dx 
\label{Int12wseg}.
\end{equation}

The $\w{\seg}_{ij}$ may be also directly associated with terms 
used in \cite{SCLD1} for decomposition of $\sigseg$ and it is revisited
below in \Sec{SigMxSeg}.

\smallskip

Finally, matrix chord length distribution $\bm{\w{\cld}}$ may be introduced
\begin{equation}
\w{\cld}_{ij}(l) = \frac{4}{S_\cup}\cor''_{ij}(l).
\label{wcld}
\end{equation}
Signed chord length distribution may be expressed for the union due to \Eq{corU} as
\begin{equation}
 \sigcld(l) = \sum_{i,j=1}^n \w{\cld}_{ij}(l).
\label{cldW}
\end{equation}
Analogue of Dirac integral for two bodies may be derived from
\Eq{Int12CorB''}
\begin{equation}
\frac{1}{V_\cup}\int_{\B_i} \int_{\B_j}%
 \frac{\varphi(|\ve{r}-\ve{r}'|)}{4 \pi |\ve{r}-\ve{r}'|^2} d\ve{r}\, d\ve{r}' 
 = \frac{S_\cup}{4V_\cup} \int_0^\infty\!\!\w{\cld}_{ij}(x)%
 \left(\int_0^x\!\!\!\int_0^p\!\!\varphi(r)dr\,dp\right) dx. 
\label{Int12wcld}
\end{equation}
For $\B_i = \B_j = \CB$ \Eq{Int12wcld} coincides with integral \EqI{2.5} used 
in {\em Dirac's method of chords}.  

Relation of $\w{\cld}_{ij}$ with decomposition used in \cite{SCLD1} for 
construction of $\sigcld$ is discussed further in \Sec{SigMxCld}.

\subsection{Radii (signed matrix) density function}
\label{Sec:SigMxSeg}

\begin{figure}[hbt]
\begin{center}
\includegraphics[scale=0.24]{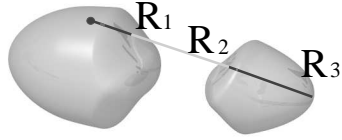}
\end{center}
\caption{Scheme of intervals for radii in two disjoint bodies}\label{Fig:B2rad}
\end{figure}

Consideration below is very similar with \cite[Section 3.2]{SCLD1}.
After transition to spherical coordinates discussed in first part
\EqI{2.8, A9} it is possible to express \Eq{Int12B4} via analogue 
of \EqI{3.2}
\begin{equation}
\int_{\B_1} \int_{\B_2}%
 \frac{\varphi(|\ve{r}-\ve{r}'|)}{4 \pi |\ve{r}-\ve{r}'|^2} d\ve{r}\, d\ve{r}' 
=\frac{1}{4 \pi }\int_{\B_1}\!\!d\ve{r}\int d\Omega\!%
\int_{\Rd \cap \B_2}\!\! \varphi(R) dR,  
\label{IntDirVSRB12}
\end{equation}
where ${\Rd \cap \B_2}$ is intersection of a body $\B_2$ with a ray 
from a point inside the body $\B_2$, \eg segment $[R_2,R_3]$ on \Fig{B2rad}.

Let's denote for a ray from $\B_i$ as $R^{(ij)}_k$ the distance between origin 
and $k$-th intersection with $\B_j$, {\eg} for two convex bodies on \Fig{B2rad}
$R^{(12)}_1=R_2$ and $R^{(12)}_2=R_3$. Origin is considered further as {\em first} 
intersection. {\ie} $R^{(11)}_1=0$ and $R^{(11)}_2=R_1$ on \Fig{B2rad}. In more 
general case with nonconvex source body $R^{(12)}_1 = R_{2m}$ with $m \ge 1$.

With such notation 
\begin{equation}
\int_{\Rd \cap \B_2}\!\! \varphi(R) dR \equiv
 \sum_{k=1}^{n^{(12)}} \int_{R_{2k-1}^{(12)}}^{R_{2k}^{(12)}}\!\! \varphi(R) dR
 = \sum_{k=1}^{2n^{(12)}} (-1)^k \int_0^{R_{k}^{(12)}}\!\! \varphi(R) dR,  
\label{SumSegB12}
\end{equation}
where $n_{12}$ is amount of intervals of ray from $\B_1$ inside body $\B_2$. 
A similar equation is appropriate for any number of bodies and any pair 
$\B_i$, $\B_j$.
It may be written formally due to \Eq{SumSegB12}
\begin{equation}
\w{\seg}_{ij}(l) = \sum_{k=1}^{2n^{(ij)}_{\rm max}} (-1)^k \seg_k^{(ij)}(l),
\label{pmSegij}
\end{equation}
where $\seg_k^{(ij)}(l)$ is density function for length of $k$-th 
intersection with $\B_j$ of ray originated in $\B_i$. 
It is an analogue of \EqI{3.5} up to sign $(-1)^{k+1}$ due to
formally zero-based indexes in initial numeration ({\eg} $R_0 = 0$) used
in \cite{SCLD1}.

It is analogy of \cite[Section 3.2]{SCLD1} with representation
of $\sigseg$ as an alternating sum of $\seg_k$ \EqI{3.2}. The main
difference of representation $\w{\seg}_{ij}$ as a sum $\seg_k$ is
requirement for origin of ray to be inside $\B_i$ and inclusion
in sum only intervals inside $\B_j$.
For a case depicted on \Fig{B2rad} $\seg_2^{(11)}(l)$, $\seg_1^{(12)}(l)$, 
and $\seg_2^{(12)}(l)$ are distributions of radii $R_1$, $R_2$, and $R_3$ 
respectively and would correspond in initial notation \cite{SCLD1} to 
$\seg_1$, $\seg_2$ and $\seg_3$. 

For nonconvex target $R^{(ij)}_1 = R_{2m}$ with different $m \ge 1$  
(if $i \ne j$) there is no direct relation between $\seg_k^{(ij)}$ in 
\Eq{pmSegij} and $\seg_{k'}$ with some fixed $k'$ in notation used in first 
part \EqI{3.2}. 
It is rather formal rearrangement, because due to \Eq{segW} together with \Eq{pmSegij}
there is an analogue of \EqI{3.5}
\begin{equation}
 \sigseg(l) = \sum_{i,j,k=1} (-1)^k \seg_k^{(ij)}(l),
\label{sumsegijk}
\end{equation}
there $\sigseg(l)$ is signed radii distribution for collection $\B_i$.
Such correspondence produce normalization for all $\w{\seg}_{ij}$. It
is not always convenient for analysis of single integral like \Eq{Int12},
\Eq{Int12wseg}, \Eq{IntDirVSRB12} {\em etc}.
  
Say, for case with two bodies $V_\cup = V_1 + V_2$ in \Eq{Int12wseg} 
it corresponds to already noted in \Eq{wsegii} scaling 
$\sigseg^{\B_1} = \w{\seg}_{11}(l)\,V_\cup/V_1$
and $\sigseg^{\B_2} = \w{\seg}_{22}(l)\,V_\cup/V_2$. 
Asymmetric normalization {\em on source body} produces yet another notation
\begin{equation}
 \seg_i^j(l) \equiv \frac{V_\cup}{V_i}\w{\seg}_{ij}(l)
\quad \Longrightarrow \quad
\seg_i^i(l) = \sigseg^{\B_i}(l),
\quad
 \seg_i^j(l) = \frac{V_j}{V_i}\, \seg_{j}^i(l)
\label{segaij}
\end{equation}
with last equation due to symmetry of initial definition
$\w\seg_{ij} = \w\seg_{ji}$.

Here is also convenient to use stochastic model similar with
introduced in \cite{SCLD1}. Each ray is ``primary event''
and for any given body $\B_i$ there are $n$ kinds of ``secondary
events'': intersection of the ray with boundary of $\B_i$ and
intersections with other ($n-1$) bodies. Each such event has
``negative sign'' if the ray enters into the body and positive
one otherwise.

For all bodies nonoverlapping with $\B_i$ number of
``negative'' and ``positive'' intersection are equal
and so formal ``balance'' $N^{(ij)} \equiv N_+^{(ij)} - N_-^{(ij)}$ 
for such events is zero.
It is yet another demonstration of zero integral 
$\int_0^\infty\seg_{ij}(l)=0$ for $i \ne j$ and disjoint 
bodies. More direct explanation is rather obvious equality 
of such integrals over corresponding ``negative'' and ``positive'' 
terms $\seg_k^{(ij)}(l)$ and $\seg_{k+1}^{(ij)}(l)$ in
\Eq{pmSegij}.

\subsection{Chord length (signed matrix) density function}
\label{Sec:SigMxCld}

\begin{figure}[hbt]
\begin{center}
\includegraphics[scale=0.24]{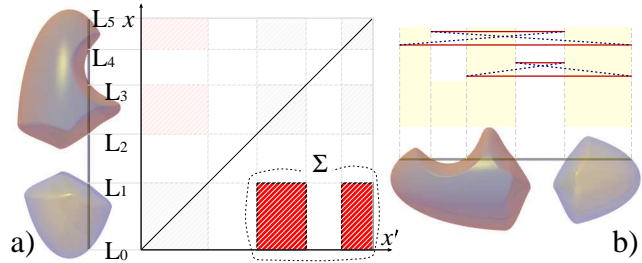}
\end{center}
\caption{a) Scheme of integration. b) Scheme of chords quadruplets}
\label{Fig:Dib12chord}
\end{figure}

This section is based on \cite[Section 3.3]{SCLD1}. 
For calculation of $\cld_{12}(l)$ it is necessary
to save in \EqI{3.14, 3.15} only terms with $x \in \B_1$, $x' \in \B_2$.
Each such term corresponds to an integration on some {\em rectangle}
$x \times x' \in [L_{2k},L_{2k+1}] \times [L_{2m},L_{2m+1}]$
represented on \Fig{Dib12chord} derived from analogous
scheme in \cite[Figure~4]{SCLD1}.

It may be written instead of \EqI{3.13}
\begin{equation}
 \int_{\B_1} \int_{\B_2}%
 \frac{\varphi(R)}{4 \pi R^2} d\ve{r}\, d\ve{r}' 
 = \frac{1}{4 \pi}\int\!\Bigl(-\!\int_{\Lh\cap\B_1}\int_{\Lh\cap\B_2}%
\!\!\varphi(x'-x)dx\,dx'\Bigr) d\T,
\label{IntDirTxxB12}
\end{equation}
there $\Lh\cap\B_k$ is intersection of line $\Lh$ with body $\B_k$ and
$d\T$ is measure of integration on space of lines used earlier in \cite{SCLD1}. 
Here area of integration $(\Lh\cap\B_1) \times (\Lh\cap\B_2)$ is union of
rectangles mentioned above and denoted as $\SQU_{k,m}$ in \EqI{3.12c}.

If to apply method developed in \cite[Section 3.3]{SCLD1} to
Dirac integral for body $\NB = \B_1 \cup \B_2$, then area of
integration considered here becomes subset of $\LLint{(\B_1 \cup \B_2)}$
used in \EqI{3.13}.

More generally, decomposition \Eq{cldW} corresponds to expressions
of Dirac integral \EqI{3.20} for $\bigcup_k\B_k$ via sum of Dirac 
integrals for $\B_k$ and integrals \Eq{IntDirTxxB12} for pairs 
$\B_i$, $\B_j$. Necessary expressions may be found in first part 
\EqI{3.14, 3.15}. 

Let's denote intersections of line with boundary of body $\B_j$ as $L^{(j)}_k$,
$k = 0, \ldots 2n_{\rm max}-1$. Then
\begin{equation}
\SQU_{k,m}^{(ij)} = \{(x,x') : L_{2k}^{(i)} \le x \le L_{2k+1}^{(i)},\,
  L_{2m}^{(j)} \le x' \le L_{2m+1}^{(j)}\}.
\label{AChrSQij}
\end{equation}
Such terms corresponds to contributions of $\w\cld_{ij}$ 
in two last sums in \EqI{3.15} via quadruplets expressed by \EqI{3.14c}.

Let us use notation \EqI{3.11}
\begin{equation}
 \TRI^L(\varphi) \equiv
 \int_0^L\!\!\int_0^p\!\! \varphi(r)dr\,dp 
\label{Iintpr}
\end{equation}
From \EqI{3.14c} rewritten with new notation \Eq{AChrSQij} follows
\begin{equation}
\begin{split}
\SQU_{k,m}^{(ij)}(\varphi)
&\equiv -\iint_{\SQU_{k,m}^{(i,j)}}\!\!\varphi(x'-x)dx\,dx'
 = -\int_{L_{2k}^{(i)}}^{L_{2k+1}^{(i)}}
\!\!\int_{L_{2m}^{(j)}}^{L_{2m+1}^{(j)}}\!\! \varphi(x'-x)dx' dx \\
&=\TRI^{L_{2m}^{(j)}-L_{2k+1}^{(i)}}(\varphi) 
- \TRI^{L_{2m}^{(j)}-L_{2k}^{(i)}}(\varphi) 
 -\TRI^{L_{2m+1}^{(j)}-L_{2k+1}^{(i)}}(\varphi) 
 + \TRI^{L_{2m+1}^{(j)}-L_{2k}^{(i)}}(\varphi).
\end{split}
\label{IntSQUij}
\end{equation}
For example, \Eq{IntDirTxxB12} may be rewritten using such notation as
\begin{equation}
 \int_{\B_1} \int_{\B_2}%
 \frac{\varphi(R)}{4 \pi R^2} d\ve{r}\, d\ve{r}' 
 = \frac{1}{4 \pi}\int d\T\sum_{k,m}\SQU_{k,m}^{(12)}(\varphi),
\label{IntTB12SQU}
\end{equation}
where $k$ and $m$ are zero-based indexes of intervals of a chord inside of
first and second body respectively, {\eg} for two convex
bodies it would be only one term and for scheme on \Fig{Dib12chord}a
there are up to two terms. 

Due to \Eq{IntSQUij} each such a term formally 
corresponds to four segments of a chord \Fig{Dib12chord}b.
So, for two different bodies each pair of intervals of intersection 
with the same line $[A,B] \in \B_i$ and $[C,D] \in \B_j$   
generates quadruplets of chord segments with ``positive'' 
pair $[AD]$, $[BC]$ and ``negative'' pair $[AC]$, $[BD]$.

Let us denote $\cld^{(ij)}_{mk}$ distribution of length 
$L^{(j)}_m - L^{(i)}_k$. It is clear from \Eq{IntSQUij} that
sign is always equal to $(-1)^{m-k+1}$ and so it is possible
to define
\begin{equation}
 \cld_{ij}(l) = \sum_{k,m} (-1)^{m-k+1} \cld^{(ij)}_{km}
\label{cldijmk}
\end{equation}

Let's also rewrite \Eq{IntTB12SQU} using mentioned rule for signs of 
intervals
\begin{equation}
 \int_{\B_i} \int_{\B_j}%
 \frac{\varphi(R)}{4 \pi R^2} d\ve{r}\, d\ve{r}' 
 = \frac{1}{4 \pi}\int (-1)^{m-k+1}\sum_{k,m}\TRI_{k,m}^{(ij)}(\varphi)\, d\T.
\label{IntTBijTRI}
\end{equation}
 
Application of methods represented in 
\cite[Sec.~2.2, Sec.~3.3, App.~A-5]{SCLD1} to \Eq{cldijmk} and \Eq{IntTBijTRI}
produces yet another version of \Eq{Int12wcld}
\begin{equation}
 \int_{\B_i} \int_{\B_j}%
 \frac{\varphi(|\ve{r}-\ve{r}'|)}{4 \pi |\ve{r}-\ve{r}'|^2} d\ve{r}\, d\ve{r}' 
 = \mathscr{S}_{\B_i}^{\B_j} \int_0^\infty\!\!\cld_{ij}(x)%
 \left(\int_0^x\!\!\!\int_0^p\!\!\varphi(r)dr\,dp\right) dx, 
\label{Int12cld}
\end{equation}
where $\mathscr{S}_{\B_i}^{\B_j}$ is constant. Due to principles 
discussed in \cite[App.~A-5]{SCLD1} it is equivalent with average 
overlap of projection of $\B_i$ and $\B_j$ on the same plane. 

Such normalization produces some difficulty, due to absence of simple
generalization of Cauchy formula for average area of surface of 
{\em single body}, {\eg} $\mathscr{S}_{\B}^{\B} = S/4$.
Average chord length \cite{SCLD1} also may not be used, because 
it is zero
\begin{equation}
 \int_0^\infty l\,\cld_{ij}(l) dl  \propto -\int_0^\infty l \seg'_{ij}(l) dl
 = \int_0^\infty \seg_{ij} dl = 0
\label{lcld0}
\end{equation}

Let's discuss that more explicitly. In \cite[App.~A-5]{SCLD1} 
for each line $L$ intersecting {\em single convex body} is obviously 
defined a chord length $\Lh(L)$. So the chord length 
$\Lh(L)$ may be considered as some function on space of lines.  
This space $\T$ may be considered as some abstract space and
Dirac ``chord integral'' may be expressed \cite[A20]{SCLD1} via 
integration on this space.

For nonconvex body with $n_I$ intervals of intersection in
\cite[Sec.~3.3]{SCLD1} similar integral was defined via
sum of $2 n_I^2 - n_I$ terms with different signs. 
Formally instead of one function $\Lh(L)$ on the space of lines $\T$
it may be considered $2 n_{\rm max}^2 - n_{\rm max}$ different 
functions $\Lh_{k,m}(L)$, $k,m = 1, \ldots, 2 n_{\rm max}$, $k < m$,
where $n_{\rm max}$ is maximal number of
intervals, $k$ and $m$ are indexes of intersections.

For each function $\Lh_{k,m}(L)$ may be defined ``density function
of interval $[k,m]$'' $\cld_{k,m}(l)$ and it is very similar with 
definition of $\Lh(L)$ for intersection with single convex body. It
justifies application of methods developed in \cite[App.~A-5]{SCLD1} 
for nonconvex case. 
In \cite[Sec.~3.3]{SCLD1} notation $\Lh_{k,m}(L)$, $\cld_{k,m}(l)$ was 
not used, because these distributions was combined for 
brevity in three groups: $\cld_1(l)$, $\cld_+(l)$, $\cld_-(l)$.

The same consideration may be used here, if to consider few bodies 
as one compound nonconvex object with two additional indexes like 
in \Eq{AChrSQij} to mark ``source'' and ``target'' bodies. 
So, there are distributions of length $\Lh_{k,m}^{(ij)}(L)$ 
used in equations like \Eq{cldijmk}.
For two bodies only combination
of indexes $12$ (or $21$) is useful for calculation of integrals
like \Eq{Int12} and all other terms are used for connection with 
theory developed in \cite{SCLD1}. 

\smallskip

It is possible again to use common normalization on union of bodies
similar with \Sec{SigMxSeg} to produce \Eq{Int12wcld},
coinciding with \Eq{Int12cld} up to constant and term $\w\cld_{ij}$.
In such a case sum \Eq{IntTBijTRI} is considered as part of
sum \EqI{3.15} in expressions for Dirac integral with single
nonconvex object constructed as the union of all $\NB = \bigcup_k \B_k$.

Area of integration denoted in \EqI{3.13} as $\LLint{\NB}$ for
each line may be represented as union of $\LLint{\B_j}$ for each
body with already considered sets $(\Lh \cap \B_i) \times (\Lh \cap \B_j)$
for each pair of bodies. It may be clarified by comparison of
\Fig{Dib12chord}a and \cite[Fig.~4a]{SCLD1}.
So signed chord distribution for union of bodies $\sigcld^\cup$ is 
constructed as sum of distributions $\w\cld^{(ij)}$ for all pairs 
of bodies like \Eq{cldW}.  

The only difference between $\cld^{(ij)}$ and $\w\cld^{(ij)}$ is
normalization. First one is normalized on set of lines intersecting
both bodies $\B_i$ and $\B_j$. Second one is normalized
on intersection with union, {\ie} at least with one body $\B_k$.
Second method produce normalization using simply defined values 
$V_\cup = \sum_k V_k$ and $S_\cup = \sum_k S_k$.

Here is again may be used stochastic model similar with discussed
at end of {\em Section 3.3} of \cite{SCLD1}. It is used uniform
isotropic set of lines, and for each chord are generated $n^2$ 
signed distributions instead of only one, because each segment
started in $\B_i$ and finished in $\B_j$ is marked by two additional 
indexes $i$ and $j$. 

Different kinds of segments due to intersection with such lines produce 
distributions of lengths $\w{\cld}_{k,m}^{(ij)}(l)$ with signed sums
equivalent to \Eq{cldijmk} up to multiplier
\begin{equation}
 \w\cld_{ij}(l) = \sum_{m,k} (-1)^{m-k+1} \w\cld^{(ij)}_{k,m}(l)
\label{wcldijmk}
\end{equation}
and so generate matrix $\bm{\w{\cld}}(l)$ representing decomposition of
$\sigcld(l)$ for $\bigcup_k\B_k$ already presented earlier \Eq{cldW}.

There are two ``positive'' and two ``negative'' segments in each 
quadruplet \Eq{IntSQUij} and ``event balance'' 
$N^{(ij)} \equiv N_+^{(ij)} - N_-^{(ij)}$ 
is zero. It is confirmation of property
$\int_0^\infty\cld_{ij}(l)=0$ for $i \ne j$ and disjoint 
bodies. Sums of lengths for two positive and two negative segments 
are equivalent 
$(L_{2m}^{(i)}-L_{2k+1}^{(j)})+(L_{2m+1}^{(i)}-L_{2k}^{(j)}) = 
(L_{2m}^{(i)}-L_{2k}^{(j)})+ (L_{2m+1}^{(i)}-L_{2k+1}^{(j)}), 
$ 
so contribution to average length is zero, {\ie} 
$\int_0^\infty l\cld_{ij}(l)=0$, {\em cf\/} \Eq{lcld0}. 

\subsection{Signed\/ $\lambda$--chords}
\label{Sec:SigLam}

Yet another way to set normalizing multiplier is to use an analogue
of ``$\lambda$-randomness'' ({\em cf} \cite{SCLD1} and references therein), 
{\ie} some function with extra multiplier 
$\lambda_{ij}(l) \propto l^4\cld_{ij}(l)$.

For such a case it is always possible to normalize $\lambda(l)$
with condition $\int_0^\infty \lambda_{ij}(l) dl = 1$
and to write instead of \Eq{Int12cld}
\begin{subequations}\label{Int12lamcld}
\begin{equation}
 \int_{\B_i} \int_{\B_j}%
 \frac{\varphi(|\ve{r}-\ve{r}'|)}{4 \pi |\ve{r}-\ve{r}'|^2} d\ve{r}\, d\ve{r}' 
 = C^{\lambda}_{ij} \int_0^\infty\!\!\frac{\lambda_{ij}(x)}{x^4}%
 \left(\int_0^x\!\!\!\int_0^p\!\!\varphi(r)dr\,dp\right) dx, 
\end{equation}
where $C^{\lambda}_{ij}$ may be simply calculated using $\varphi(r) = 4 \pi r^2$
\begin{equation}
 \int_{\B_i} \int_{\B_j} d\ve{r}\, d\ve{r}' = V_i V_j  =
C^{\lambda}_{ij} \int_0^\infty\!\!\frac{\pi}{3}\lambda_{ij}(x)dx
\quad \Longrightarrow \quad 
C^{\lambda}_{ij} = \frac{3}{\pi} V_i V_j.
\end{equation}
\end{subequations}

Unlike single convex body here $\lambda$-chords formally may {\em not} 
be defined by lines through pair of points with independent uniform 
distributions inside body. Here $\lambda_{ij}(l)$ rather should be 
considered as some formal function produced from $l^4\cld_{ij}(l)$ or 
$l^4\cor''_{ij}(l)$ after normalization on unit. Yet, for $i=j$ and convex $\B_i$ 
it may be derived from initial definition.

Another equation may be produced directly from comparison of
\Eq{Int12CorB''} and \Eq{Int12lamcld}
\begin{equation}
 C^{\lambda}_{ij}\frac{\lambda_{ij}(x)}{x^4} = \cor''_{ij}(x)
\quad \Longrightarrow \quad 
\lambda_{ij}(x) = \frac{\pi}{3}\frac{x^4\cor''_{ij}(x)}{V_i V_j}.
\label{cor2lam}
\end{equation}

Let's express $\w{\cld}_{12}(l)$ from $\lambda_{12}(l)$ for
two bodies using \Eq{wcld} and \Eq{cor2lam}
\begin{equation}
 \w{\cld}_{12}(l) = \frac{4}{S_1+S_2} \cor''_{12}(l)
 = \frac{12 V_1 V_2}{\pi (S_1+S_2)} \frac{\lambda_{12}(l)}{l^4}.
\label{lam2wcld}
\end{equation}

\section{Nonuniform case}
\label{Sec:Nonun}

For nonuniform case for two bodies with densities $\rho_1$ and $\rho_2$ instead 
of \Eq{Int12} or \Eq{Int12B4} may be written an analogue of \EqI{B1}
\begin{eqnarray}
 \iint \rho_1(\ve{r}) \rho_2(\ve{r}')%
 \frac{\varphi\bigl(|\ve{r}'-\ve{r}|\bigr)}{4\pi|\ve{r}'-\ve{r}|^2} d\ve{r}\, d\ve{r}' 
 &=& \int_0^\infty \cor_{12}(x) \varphi(x)  dx \nonumber \\
 &=& \int_0^\infty\!\!\cor_{12}''(x)%
 \left(\int_0^x\!\!\!\int_0^p\!\!\varphi(r)dr dp\right) dx.
\label{Int12Mcor}
\end{eqnarray}
It follows from definition of correlation function \Eq{corr2} together
with \Eq{Int12C4} and integration by parts, {\em cf} \Eq{Int12B4}.

For nonoverlapping $\rho_1$ and $\rho_2$: $\cor_{12}(0) = \cor'_{12}(0) = 0$ 
and there is problem with normalization, {\em cf} \EqI{B2}. It is
useful to define some $\acute{\lambda}(l) \propto l^4\cor''(l)$
and to write analogue of \Eq{Int12lamcld}
\begin{subequations}\label{Int12lamrho}
\begin{equation}
 \iint \rho_1(\ve{r}) \rho_2(\ve{r}')%
 \frac{\varphi(|\ve{r}-\ve{r}'|)}{4 \pi |\ve{r}-\ve{r}'|^2} d\ve{r}\, d\ve{r}' 
 = \acute{C}^{\lambda}_{12} \int_0^\infty\!\!\frac{\acute{\lambda}_{12}(x)}{x^4}%
 \left(\int_0^x\!\!\!\int_0^p\!\!\varphi(r)dr\,dp\right) dx, 
\end{equation}
where $C^{\lambda}_{12}$ may be again calculated using $\varphi(r) = 4 \pi r^2$
\begin{equation}
 \iint \rho_1(\ve{r}) \rho_2(\ve{r}') d\ve{r}\, d\ve{r}'  = M_1 M_2  =
\acute{C}^{\lambda}_{12} \int_0^\infty\!\!\frac{\pi}{3}\acute{\lambda}_{12}(x)dx
\quad \Longrightarrow \quad 
\acute{C}^{\lambda}_{12} = \frac{3}{\pi} M_1 M_2,
\end{equation}
\end{subequations}
where $M_1 = \int\rho_1(\ve{r})d\ve{r}$ and  $M_2 = \int\rho_2(\ve{r}')d\ve{r}'$
are masses of the bodies. So, $\acute{\lambda}_{ij}(x)$ may be formally defined 
by equation similar with \Eq{cor2lam}
\begin{equation}
 \acute{C}^{\lambda}_{ij}\frac{\acute{\lambda}_{ij}(x)}{x^4} = \cor''_{ij}(x)
\quad \Longrightarrow \quad 
\acute{\lambda}_{ij}(x) = \frac{\pi}{3}\frac{x^4\cor''_{ij}(x)}{M_i M_j}.
\label{aclam}
\end{equation}

\section{Arbitrary paths}
\label{Sec:Paths}

\begin{figure}[hbt]
\begin{center}
\includegraphics[scale=0.75]{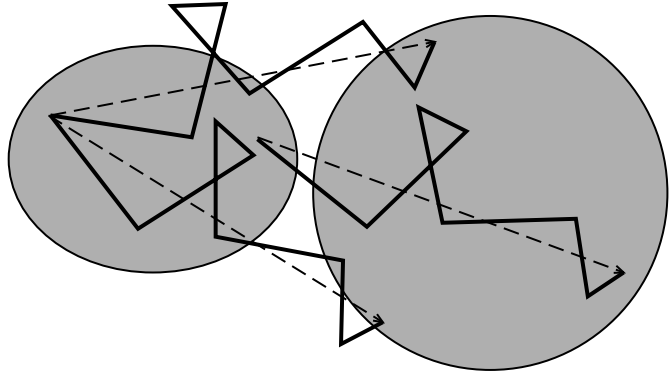}
\end{center}
\caption{Different paths with isotropic distribution}
\label{Fig:paths12}
\end{figure}

Discussion in \cite[Sec.~5]{SCLD1} about possibility of applications to arbitrary 
paths with uniform and isotropic distribution of initial points and directions  
is certainly true for presented consideration with few bodies, see \Fig{paths12}. 
The ``ray-tracing'' schemes like \Fig{transf12}, \Fig{B2rad}, 
\Fig{Dib12chord}, {\em etc.} have rather illustrative purposes and 
the only necessary condition --- is spherical and translational symmetry 
due to argument $R=|\ve{r}-\ve{r}'|$ of function $\Phi(R)$ in \Eq{Int12} 
and other expressions.

\begin{thebibliography}{9}
\bibitem[{\sf Part I}]{SCLD1} A. Yu. Vlasov, \tit{Signed chord length distribution. I}
Preprint arXiv:0711.4734 [math-ph] (2007) and references therein.
\end{thebibliography}
\end{document}